# Photonic integrated beam delivery for a rubidium 3D magneto-optical trap

Andrei Isichenko[1], Nitesh Chauhan[1], Debapam Bose[1], Jiawei Wang[1], Paul D. Kunz[2], and Daniel J. Blumenthal[1*]

[1] Department of Electrical and Computer Engineering, University of California Santa Barbara, Santa Barbara, CA 93106 USA

[2] DEVCOM U.S. Army Research Laboratory, Adelphi, MD 20783 USA

* Corresponding author (danb@ucsb.edu)

## ABSTRACT

Cold atoms are important for precision atomic applications including timekeeping and sensing. The 3D magneto-optical trap (3D-MOT), used to produce cold atoms, will benefit from photonic integration to improve reliability and reduce size, weight, and cost. These traps require the delivery of multiple, large area, collimated laser beams to an atomic vacuum cell. Yet, to date, beam delivery using an integrated waveguide approach has remained elusive. We report the demonstration of a $^{87}$Rb 3D-MOT using a fiber-coupled photonic integrated circuit to deliver all beams to cool and trap > 1 x $10^6$ atoms to near 200 μK. The silicon nitride photonic circuit transforms fiber-coupled 780 nm cooling and repump light via waveguides to three mm-width non-diverging free-space cooling and repump beams directly to the rubidium cell. This planar, CMOS foundry-compatible integrated beam delivery is compatible with other components, such as lasers and modulators, promising system-on-chip solutions for cold atom applications.

# INTRODUCTION

Cold atoms[1–4] are a central component of precision scientific tools including atomic clocks[5–8] and ultra-high resolution spectroscopy[9–11] and are an important technology for applications such as atomic timekeeping[12], quantum computing[13–18], and quantum sensing[19–21] by enabling improved spectroscopy resolution for observing the quantum behavior of atoms[3]. Today's atom trapping and cooling systems employ free-space lasers and optics that occupy tables and racks and are costly and power consuming. Realizing these systems at the chip-scale, with a complementary oxide semiconductor (CMOS) foundry-compatible photonic integrated circuit (PICs) process will improve their reliability, lower their cost, and enable portability[19,22] and new cold-atom systems for applications such as mobile gravity mapping[20] and space-based atomic clocks[23].

The three-dimensional (3D) magneto-optical trap (MOT)[1–4, 24–29], in particular, is a widely used configuration to directly create a large population of trapped and cooled atoms. The trap is formed inside a vacuum environment using a balance of optical radiation forces via six cooling beams and a magnetic field gradient with a null located at the beam intersection of the cooling beams[24]. As the trapped atom number scales strongly with the beam diameter, beam diameters of at least 1 mm are needed to trap and cool over 1 million cold atoms as is desired for improved application sensitivity[30,31]. Efforts to miniaturize[31] the beam delivery components in a MOT have primarily focused on micro-optic reflective components such as pyramidal mirrors[32,33] or diffraction gratings[34–38] to convert a collimated free-space input beam into multiple intersecting beams. Diffraction grating MOTs (GMOTs) have emerged as a robust candidate and have recently been demonstrated in a compact sensor for atom interferometry[21]. Further MOT size reduction efforts employed micro-fabricated pyramids etched into silicon[39], but have the tradeoff of a small beam overlap volume that limits the atom number. Free-space illuminated meta-surface structures capable of high efficiency MOT beam delivery[40] utilize bulk optic components for beam expansion and involve complex electron-beam lithography during fabrication. These approaches require bulk free-space optic components for beam expansion and the distance required for expansion optics poses a limit to the minimum size of the MOT package. Photonic integrated circuit (PIC) technology in the silicon nitride ($Si_3N_4$) platform[41] offers a rich set of passive and active components with low optical losses and compatibility with CMOS foundry processes at cold atom visible wavelengths[41–43]. To date, $Si_3N_4$ PICs have been used to demonstrate transformation of 780 nm light from a waveguide to a free-space beam via an on-chip surface grating, producing a free-space beam diameter of 320 um[44] and 3 mm[45] and used as a probe beam in a rubidium spectroscopy reference[46] and micrometer-size focused beams for ion-trap quantum computing[47]. Recently, a meta-surface beam expander above a single PIC emitter was used to generate a diverging, cm-size beam as the input for a GMOT[37]. However, to date, 3D-MOT atom cooling using an integrated waveguide to free-space beam delivery system has not been reported.

In this work we demonstrate for the first time, to the best of our knowledge, laser cooling and trapping of $^{87}Rb$ atoms in a 3D-MOT using a photonic integrated fiber-coupled photonic chip for the delivery of cooling and repump beams directly to a rubidium vapour cell. This photonic integrated circuit (PIC) enables a 3D-PICMOT, based on a CMOS foundry compatible $Si_3N_4$ waveguide circuit[41] for fiber coupling, beam expansion, collimation, and delivery to the vapour cell. The PIC transforms fiber-coupled 780 nm cooling and repump light into three 2.5 mm x 3.5 mm ($e^{-2}$ diameter) collimated free-space beams[45] producing a beam overlap volume of approximately 22 $mm^3$ inside the vapour cell at a height 9 mm above the PIC surface. Using this PIC delivery interface, we measure a MOT cloud of ~1.3 x $10^6$ atoms and a temperature near 200 μK using time of flight (TOF) temperature measurements. The PIC supports delivery of intensities of over 3 $I_{sat}$ (where $I_{sat}$ = 3.6 mW cm$^{-2}$ for the $^{87}$Rb $D_2$ transition) per cooling beam owing to the low absorption and the high-power handling capability of the $Si_3N_4$ waveguides. The power loss from the fiber input to the sum of the free-space emitted beams is 16 dB, which includes the loss from the packaged fiber input, excess loss of the 1x3 splitter, waveguide and mode transformer, and grating losses. Strategies to reduce this loss below 8 dB are discussed. This integrated beam delivery approach can be designed to support wavelengths from 405 nm – 2350 nm[41] and can be integrated with other key active and passive components[48–51], including lasers[52,53], ultra-low power consumption modulators[51], and reference cavities[54], to further reduce the size of 3D-MOTs. The system can be designed for operation at wavelengths for different atom species including, for example, neutral strontium (atomic clock transitions in the range from 461nm - 813 nm[6]) and cesium (MOTs at 852 nm[55]). These results show promise for the next generation of ultra-compact and portable cold atom systems.

## RESULTS

### PICMOT system design and assembly

A diagram of the laser cooling beam delivery PIC configured in a 3D-MOT is shown schematically in Figure 1. The MOT laser light is generated with two independent lasers, cooling (780.241 nm) and repump (780.228 nm), combined using a fiber directional coupler, and packaged using epoxy to the PIC input waveguide. Two current-carrying anti-Helmholtz coils are aligned along one beam axis as shown to form a quadrupole magnetic field with a null at the beam intersection within the vapour cell. The quarter waveplates (QWPs) are mounted directly above the chip and are rotated such that the circular polarization handedness of beams B1 and B3 is opposite that of B2. Three retroreflectors, each containing a mirror and quarter waveplate, are used to produce the other three of the six MOT trapping beams with the correct polarization states.

The PIC delivers non-diverging cooling and repump beams that intersect at a height of 9.45 mm above the PIC surface. This allows the PIC to be located outside of the glass vacuum cell and provide a compact configuration. The cell containing the rubidium atoms is a ColdQuanta™ rubidium miniMOT. A key novel feature of our design that makes the MOT more compact, is that the PIC delivers all beams through the same glass cell wall, which is different than typical 3D-MOTs that deliver one beam per glass wall. To reduce stray reflections and higher order modes, we use a matte-black aluminum foil baffle layer above the quarter wave plates. The retroreflector mirrors are aligned above the cell to form overlapping counter-propagating beams. The magnetic field is set by adjusting the relative current in each coil and with external magnets for field-shimming.

### Laser cooling and repump beam delivery PIC

The 3D-MOT beam delivery PIC is based on a low-loss $Si_3N_4$ core and $SiO_2$ cladding single mode waveguide designed to operate at 780 nm (cross section shown in Supplementary Figure 2(a)). The cooling and repump laser light is coupled via a single mode optical fiber to a single mode waveguide at the PIC input as shown in Fig. 2. The fiber-coupled input is waveguided to a 1 x 3 multimode interference (MMI) waveguide splitter with each output routed to one of three slab waveguide beam expanders, each uniformly illuminating one of three large-area surface grating emitters[45]. The grating emitter free-space collimated beams are designed to cool a large volume of rubidium atoms as well as provide a good beam intensity uniformity. The grating centers are positioned on a 13.5 mm diameter circle. Each emits a collimated beam at an angle of 57° from the PIC surface normal (Fig. 2(a)) towards the circle center, producing a 93° intersection between all three beams at a point 9.45 mm above the chip surface. The output of each grating emitter is linearly polarized. Individual 10 mm diameter quarter waveplates are located directly above each grating emitter to convert to the circular polarization required for generating the MOT (see Methods). A photograph of the PIC fiber-coupled with red laser light is shown in Fig. 2(b).

The PIC emits collimated free-space beams with a large cross-sectional area of 8.75 mm$^2$. This corresponds to a record large on-chip optical mode expansion factor of 20 x $10^6$ from the 0.44 μm$^2$ area waveguided mode delivered to an operational 3D-MOT. We use a CMOS camera to image the beams incident on a screen located above the PIC surface (Fig. 2(c)). From the beam intensity cross section at a distance 5 mm away from the PIC emitter the measured beam width (defined as the $1/e^2$ diameter) dimensions are 2.5 mm by 3.5 mm (Fig. 2(d)) which corresponds to a trapping beam overlap volume of 17.3 mm$^3$. By measuring the beam widths at different distances above the PIC surface we calculate divergence angles 0.16° and 0.35° for the $x$ and $y$ axes, respectively. This corresponds to $M^2$ values of 10 and 15 in the $x$ and $y$ axes, respectively, characteristic of flat-top beams. It has been shown that trapping beams of a more uniform, flat-top intensity can achieve higher MOT atom numbers than gaussian beams due to the increased optical forces near the edges of the beam overlap volume[37]. We measure root-mean-squared (RMS) intensity variations to be <9% in the central 20% of the mode area and 12% in the central 80% of the mode area. Furthermore, the effective Rayleigh lengths ($z_R/M^2$) are 125 cm and 113 cm and the beam dimension aspect ratio changes by 20% over 35 cm (Supplementary Fig. 3). This good collimation enables placing the grating emitters further apart on the PIC layout to achieve a larger beam intersection height if necessary. The large beam size is possible due to the combination of the large width and low loss of the slab expander and the shallow etch of the grating emitter. Details of the design and fabrication of the waveguides, beam expanders, and grating emitters are given in the Methods section and Supplementary Note 3.

The average power loss from the fiber input to each free-space beam output is 21.4 dB, corresponding to a total optical loss of beam delivery (fiber input to the sum of the beams) of 16 dB. These losses include the fiber to waveguide, splitter, and waveguide propagation losses. We discuss paths forward to reduce this loss factor to below 8 dB in Supplementary Note 4. The beams (labeled B1, B2, B3) have a relative power of 28%, 28%, 44% of the input power, where this power splitting ratio does not noticeably change over time. For an input fiber power of 65 mW the PIC delivers an average power of 0.46 mW per cooling beam, corresponding to a beam intensity of 5.4 mW cm$^{-2}$, or 1.5 $I_{\text{sat}}$ (for $I_{\text{sat}} = 3.6$ mW cm$^{-2}$). A summary of the loss contributions from the PIC is in Supplementary Table 1.

## Rubidium atom cooling demonstration

We demonstrate an operational $^{87}$Rb 3D-MOT using this interface. Laser cooling light at 780.24 nm is prepared by stabilizing the cooling laser relative to a hyperfine transition on the $^{87}$Rb $D_2$ line using an external saturation absorption spectroscopy setup and a separate repumping laser is aligned to the $F = 1 \rightarrow F' = 2$ transition[25] (see Methods). The cooling laser detuning $\Delta$ is controlled with an acousto-optical frequency shifter (AOFS). The axial magnetic field gradient is 20 G/cm through the trap center.

The resulting cloud of trapped $^{87}$Rb atoms is shown in Figure 3(a). The beams emitted from the PIC fluoresce as they propagate through the vacuum cell. The inset in Fig. 3(a) shows a close-up image of the MOT cloud. From images where the camera pixels in the region of the cloud are not saturated, we estimate a cloud $e^{-2}$ diameter of 0.48 mm. The steady-state population of the MOT is measured by collecting the cloud fluorescence onto a photodiode (see Methods). With an input power of 180 mW into the PIC fiber input we achieve an atom number of $(1.3 \pm 0.2)$ x $10^6$ with the cooling laser detuned by $\Delta = -2.4\ \Gamma$. The exponential time constant for the MOT loading is measured to be 0.3 s by fitting to data (see Supplementary Fig. 3). A video of the PICMOT is shown in Supplementary Movie 1 and demonstrates the effect of changing the input light polarization and the magnetic field on the cloud.

The temperature of the trapped atom cloud is evaluated with the widely-used absorption imaging time-of-flight (TOF)[56,57] technique. We also employ an alternate release-and-recapture (RR) method to independently validate the results[58]. In both measurements, the cooling and repump beams are turned off in <5 μs (see Methods). The temperature is extracted based on the free-expansion of the cloud and assuming a Maxwell-Boltzmann velocity distribution of the atoms[56]. For TOF, a free-space beam resonant with the $^{87}$Rb cooling transition was externally routed to probe the MOT and the shut-off of the optical and magnetic fields were synchronized. The cloud widths were extracted from pairs of absorption images $I_1$ and $I_2$, where $I_1$ is the image of the probe flashed (duration 0.5 ms) at a time $t_{TOF}$ after the MOT shut-off and $I_2$ is recorded at $t_{\mathrm{TOF}} + 70$ ms when the MOT has dissipated. The optical depth (OD) is extracted using the equation $\mathrm{OD} = -\ln(I_1/I_2)$ and a Gaussian fit along each axis is taken to extract the width of the cloud. By measuring flight times $t_{TOF}$ up to 1.25 ms, we extract temperature of 183 $\pm$15 μK and 223 $\pm$18 μK in the $y$ and $z$ axes, respectively. The RR method resulted in a conservative temperature of 400$\pm$200 μK which serves as an upper bound of the cloud temperature due to an overestimate in defining the boundary of where the atoms can be recaptured. The measured temperature is close to the $^{87}$Rb $D_2$ line Doppler cooling limit of 146 μK[59] and the different temperature for each axis in each axis is due to the imbalance of forces in the MOT and different magnetic field shimming conditions. Lower, sub-Doppler-limit MOT temperatures could be achieved with additional cooling stages such as polarization-gradient cooling.

## DISCUSSION

Silicon nitride photonic integration[41] has the potential to go beyond delivery of the trapping beams, to include functions such as tunable lasers[52], modulators[51], laser stabilization[53,54], spectroscopy[46,60], and planar meta-surface retroreflectors[61]. These higher-level integrated systems-on-chip can be used to improve reliability, scalability, and size weight and power (SWaP). For example, precise and active beam power balancing can be achieved using thermal tuning $Si_3N_4$ phase shifters[62] or ultra-low-power stress-optic piezo actuators[51] to adjust the 1x3 splitter ratio. Reducing the loss from the input fiber to the PIC-delivered beams is an important next step to reduce required source laser power or enable delivery of higher power beams. Further, the PIC losses can be lowered through inverse design optimization of individual components[63] including the input fiber-to-waveguide taper, three-way splitter, slab mode expander, and beam emitter. We estimate that with the results reported in literature[64] and with recent PIC test structures we have fabricated, we can reduce the total beam delivery loss to below 8 dB (see Supplementary Note 4). This level of performance eliminates the need for expensive optical amplifiers, allowing for total cooling beam intensity delivered to the atoms to reach the $I_{\mathrm{sat}}$ for $^{87}$Rb for an input fiber-coupled power of 1-2 mW which can be produced directly with a commercial single-frequency diode laser. The achievable peak atom number in the current PICMOT may be higher than measured due to multiple variable factors which include the atomic source and vacuum system, cooling laser frequency locking, magnetic field shimming, and the repump beam delivery power. Determining the maximum achievable peak atom number for this particular PIC-based beam delivery architecture is a subject of future work.

Our PICMOT system has a modular and robust design that can be used for a variety of cold atom experiments. For example, the large beam intersection height allows for the PIC to be held *ex vacuo* enabling use with different vacuum cells designs and eliminating breaking of vacuum to swap the PIC. Due to the large silicon nitride transparency window of 405 nm – 2350 nm[49], the PIC can be readily adapted for other visible wavelengths by changing the waveguide width in the nitride patterning lithography mask, opening the potential to cool other atomic species such as cesium and strontium atoms[6]. Prior work has demonstrated grating emitters with a 90 degree angle from the PIC surface[46] and our grating design can be modified to vertically probe the trapped atom cloud for spectroscopy, atom interferometry, or an optical atomic clock. Generating multiple free-space beams from a single PIC provides an inherent stability of the optical paths which has been demonstrated to improve vibration tolerance in trapped-ion quantum computing[47]. These advantages can be beneficial in compact cold atom interferometer sensors such as for field-deployed gravity mapping measurements[20]. The PIC could deliver both the MOT and Raman probe beams and the uniform Raman beam spatial profile can improve the resulting interferometer sensitivity[65].

The atom number in 3D-MOTs is related to the trapping beam overlap volume which is dependent on the MOT configuration and the beam delivery method. Multiple studies have been done that relate the peak number of trapped atoms to the beam overlap volume such as $N \propto V^2$ for a micro-pyramid MOT[39], $N \propto V^{1.2}$ for a GMOT[35], and an experimentally determined scaling law for conventional 6-beam free-space MOTs[30]. Since precision applications such as atom interferometry[30] require a million atoms or more, GMOT designs require a 45 mm$^3$ minimum overlap volume[31]. In our PIC-based 3D-MOT (PICMOT) the estimated beam overlap volume is 22 mm$^3$ which is approximately 2 times smaller than the volume of a comparable atom number GMOT that uses a 20 mm input free-space beam diameter [21,35] (see Supplementary Figure 8). We observe that our PICMOT atom number correlates well with the conventional 6-beam free-space MOT where the beam sizes are comparable to the stronger volumetric scaling law for beam comparable to our experiment[30]. This suggests that the size of the atomic vacuum cell may be reduced without a reduction in atom number if the atomic source and vacuum properties remain constant. Our simulations show that for our existing slab beam expander, we can achieve output beam widths over 6 mm, enabling greater atom numbers. A comparison of different integrated MOT designs, their beam delivery platforms, and their performance is shown in Table 1. For techniques that achieve practical atom numbers of over 1 million, a constraint on the size of the cold atom package is the distance required for the beam expansion optics between the optical fiber input and the trap center. For example, GMOT demonstrations have used an input free-space beam with ~20 mm diameter beam using a commercial fiber collimator and bulk-optic beam expander, taking up a length of over 10 cm. Similarly, while a meta-surface beam expander can be connected above a PIC, the propagation distance of 15 cm[37] to achieve the large beam size similarly limits package volume. While the 4-beam tetrahedral configuration of a GMOT does not require retroreflectors, the beam delivery PIC has advantages in integration of other waveguide-based components that can reduce system SWaP. Table 1 also shows a comparison of optical access for different MOT platforms, defined as the number of sides available for other optical functions without blocking trapping beams. For example, the input vertical beam in a GMOT limits access to probing and imaging the MOT cloud from above. The PICMOT in a six-beam configuration has four optical access locations as well as a possible fifth if other probe beam emitters are located on the trapping beam emitter chip. A future version of the PICMOT system will have a smaller, custom-sized atomic cell which will enable bringing a more compact dome-like retroflector layer closer to the MOT. We are also actively investigating planar metasurface retro-reflectors, which can further reduce the overall PICMOT volume. More details about this comparison are discussed in Supplementary Note 6.

Future versions of the PICMOT system can introduce higher levels of system integration and active functionality, reduced optical losses, and compact magnetic field generation[38]. For example, by extending the grating emitter design to two dimensions one can achieve control of the output beam polarization to eliminate quaterwaveplates[66]. Thermal or stress-optic tuners and phase can be incorporated to control beam parameters to maximize MOT atom numbers which can be compatible with on-line multi-parameter optimization algorithms[67]. The CMOS foundry compatible fabrication process can be combined with heterogeneous integration with on-chip laser and detector technology[41]. Further integration with a narrow linewidth laser can enable compact MOTs in which all the laser delivery, interrogation, and probing is combined into a PIC. An example integrated atomic-photonic system for cold atoms is shown in Figure 4. The laser sources, reference cavities for laser stabilization[54], modulators[51], and free-space beam outputs can be co-located on a single silicon nitride PIC. By combining on-chip modulation and filtering[48], the probe beam can be shuttered and controllably detuned[60] to miniaturize the MOT cloud atom number and temperature measurements which are common cold atom sensor readout steps. This higher level of integration can facilitate the transition of atomic systems out of conventional laboratories to portable systems as well as lower the cost of cold atom experiments.

# TABLES

**Table 1.** Summary of integrated beam delivery methods for 3D MOT cold atom systems.

| Beam delivery method | Beam delivery | Beam properties | Trapping beam overlap volume | Beam delivery distance[1] | Number of retroreflectors | Optical access[2] | Atom number |
|---|---|---|---|---|---|---|---|
| Conventional MOT[31] | Six free-space beams | Collimated Gaussian beams, diameters ~30mm | ~$10^4$ $mm^3$ | N/A | 3 | 5 | ~$10^8$ |
| Diffraction grating MOT[35] | Fiber input, one collimated beam and 3 diffracted beams | Input FS beam 20 mm diameter | $10^3$ $mm^3$ | 16 cm | 0 | 3-4 | 6x$10^7$ |
| 90° Pyramid[39] | Fiber input, one collimated beam and 5 reflected beams | Input FS beam 4.2 mm diameter | 10 $mm^3$ | 3 cm | 0 | 0 | $10^4$ |
| PIC emitter, MS, GMOT[37] | Fiber input, one PIC-delivered beam and 3 diffracted beams | Diverging beam after MS. Diameter 20 mm, flat-top | 340 $mm^3$ | 15 cm | 0 | 3 | 3x$10^6$ |
| SiN beam delivery PIC (this work) | Fiber input, three PIC-delivered beams | Collimated, 2.5 x 3.5 $mm^2$, uniform intensity | 22 $mm^3$ | 4 cm | 3 | 4-5 | 1.3x$10^6$ |

[1]Approximate distance for beam expansion and routing between a single fiber-coupled input and the trap center.
[2] Number of sides available for optical access without blocking trapping beams, assuming a cube-like vacuum cell with the atomic vacuum source located at one face of the cell.

# METHODS

## PIC design and modeling

The 780 nm single mode waveguide design is a 400 nm wide, 120 nm thick $Si_3N_4$ waveguide core with a 15 µm thick thermal oxide lower cladding layer and 3 um thick PECVD oxide upper cladding layer. The waveguide cross section is shown in Supplementary Fig. 2(a). The 1 x 3 multi-mode-interference (MMI) device splits the guided 780 nm light into three single mode waveguides each coupled to a long 2D mode expander that increases the guided mode lateral dimension to 4 mm. Each 2D mode expander is terminated into an ultra-large area grating emitter of dimensions 4 mm x 5 mm to create the cooling beams.

The waveguide is designed to support a single TE0 mode, have low loss, have bend radius <500um and provide a short slab expander length while maintaining low losses. To achieve the required beam sizes for the MOT, we introduce a waveguide-to-slab mode transition within our chip. The slab expander is a wide slab of $Si_3N_4$ which allows the mode to expand freely inside it. Different waveguide core thicknesses result in different length of beam expander due to different effective refractive index of mode. The 120 nm thick $Si_3N_4$ core slab design is chosen to match the single mode waveguide design, maintain low losses, and realize a compact mode expander. For lower losses a thinner core is preferred[43] but that results in larger bend radii and longer slab expander. A thinner core results in very long tapers which are impractical, whereas a thicker core helps in making the beam expander more compact due to higher index contrast. In addition, the bend radius is also considered as thinner cores have larger minimum bend radii resulting in larger devices. For our devices, we define minimum bend radius as the radius below which the loss contribution from the bend exceeds 0.01 dB $m^{-1}$ and we ensure to keep our bend radii larger than the minimum bend radius for a core geometry. Commercially available simulation software was used to model the expansion of beam inside the taper for different core thicknesses. The simulation was performed for core thicknesses 40 nm and 120 nm with waveguides designed for each core thickness entering a taper of identical length and width. The waveguides for the 40 nm and 120 nm core width were selected to support single TE mode. The mode was propagated using beam propagation method (BPM) to observe the mode diameter expansion in the slab. From simulation, the length of the slab required for a 4 mm beam width is extremely long (>20 mm) for a 40 nm core. The corresponding minimum length of slab required for 120 nm core is 4.2 mm (see Supplementary Fig. 2 (e)). A 120 nm core is chosen as it provides a compact slab expander and a bend (critical bend radius <500 um) that makes it possible to fit a single grating in one reticle while also having low 0.3 dB $cm^{-1}$ loss. Based on results from a previous fabrication run we re-used a photolithography mask where the slab expander length was 11.43 mm. This length fits easily in a 21 mm x 25 mm reticle for our current mask without requiring stitching at the slab or grating section.

A side view diagram of the mode expander and grating emitter is shown in Supplementary Fig. 2 (b). The large beam size achieved is due to the large beam expansion in the $x$ and $y$ axes due to the design and performance of the slab expander and the grating emitter, respectively. The slab expander must be sufficiently wide such that the optical mode does not interact with the slab sidewalls and long enough to expand the mode to reach the target beam width. The large size of this photonic structure is made possible by the low optical losses in our SiN platform and this design can be scaled to larger beam expansion. The grating is etched into the $Si_3N_4$ slab waveguide core, where the 10 nm partial etch depth was designed to provide the 54.7° diffraction angle and high power in the desired diffraction order. The combination of the shallow, controllable grating etch in the relatively thick 120 nm core enables a large beam size in the other axis. The chirped apodized grating is designed with a lens-like phase curvature profile to produce uniform beam intensity[45]. The grating coupling strength is lowest at the grating input and progressively increases as the light propagates in the grating region. The grating period and duty cycle are designed to range from 1.18 um to 1.08 um and 10% to 50% from front to end of the grating respectively. For more information about the grating design choices see Supplementary Note 3.

## Fabrication process

The fabrication process (Supplementary Fig. 1) is CMOS foundry compatible[41] and starts with a 1 mm thick 100 mm (4-inch) diameter silicon wafer with 15 μm of thermal oxide. The waveguide core layer is made by depositing 120 nm of $Si_3N_4$ using Inductively Coupled Plasma-Plasma Enhanced Chemical Vapour Deposition (ICP-PECVD) at UCSB. The waveguide layer is first patterned using a standard 248 nm deep ultraviolet (DUV) stepper. The silicon nitride is then etched with an Inductively Coupled Plasma-Reactive Ion Etcher (ICP-RIE) using an optimized $CHF_3/CF_4/O_2$ etch chemistry developed at UCSB[69]. The wafer is then thoroughly cleaned and the process is repeated for the grating layer using a lower RF power during the ICP-RIE etch. The target etch depth for these gratings is 10 nm. After a final clean, ICP-PECVD is used to deposit 3 μm of $SiO_2$ upper cladding. The wafer is diced to access facets for fiber edge-coupling.

### MOT demonstration

The beam delivery PIC is bonded to an aluminum plate and fiber packaged. Conventional zero-order quartz quarter waveplates are placed in a plastic holder layer directly above the PIC and rotated to achieve the required handedness and ellipticity of polarization. The PIC setup is mounted on a multi-axis stage and aligned with respect to a pair of anti-Helmholtz coils. A commercially available vacuum vapour cell containing a rubidium dispenser (the ColdQuanta™ rubidium miniMOT) is brought into the setup and mounted directly above the PIC. Two separate lasers were used for the cooling demonstration. The cooling laser is a Distributed Bragg Reflector (DBR) 780 nm laser (from Photodigm™) and the repump is derived from frequency-doubling a 1560 nm external cavity diode laser. The cooling laser is locked to the $^{87}$Rb $5S_{1/2}, F = 2 \rightarrow 5P_{3/2}, F' = (1,3)$ cross-over transition and an acousto-optical frequency shifter (AOFS) is used to shift the cooling beam to a red detuning of 10-20 MHz relative to the $F = 2 \rightarrow F' = 3$ hyperfine transition (see Supplementary Fig. 3). Both beams are combined with a fiber-based coupler and amplified using a booster optical amplifier (BOA, Thorlabs BOA785S) for PIC input powers up to 65 mW. The BOA is used for MOT operation during the TOF measurements because of its ability to shutter the optical output with a switch-off time of less than 5 μs. For atom number and loading rate measurements, a fiber-coupled tapered amplifier (Newport™) providing a power into PIC of 180 mW is used and the fluorescence of the MOT cloud is focused onto a photodetector. The MOT temperature is measured with absorption imaging by using a separate free-space probe beam that near-resonant with the $F = 2 \rightarrow F' = 3$ generated using another AOFS. During the TOF expansion and imaging, the PICMOT magnetic field and the cooling and repumping light is turned off. There were no additional cooling stages in this MOT demonstration. Additional details of the MOT characterization are described in Supplementary Note 5.

## DATA AVAILABILITY

The data that support the plots within this paper and other finding of this study are available from the corresponding author on reasonable request.

# COMPETING INTERESTS

The authors declare no competing interests.

# AUTHOR CONTRIBUTIONS

A.I., N. C., and D. J. B. prepared the manuscript. N. C. and D. J. B. contributed to the laser cooling PIC designs and N.C. simulated the photonic structures. D. B. performed the laser cooling PIC fabrication and J.W. diced the chips. A.I. and N.C. performed the PIC beam profile measurements and built the laser stabilization system. A.I. performed the PICMOT assembly, cooling demonstration, and the MOT characterization. A.I. analyzed data to extract the MOT atom number and temperature. D. J. B and P. D. K. supervised and led the scientific collaboration.

# ACKNOWLEDGEMENTS

This work is supported by DARPA MTO APhi contract number FA9453-19-C-0030, ColdQuanta, and the UCSB Faculty Research Grant. The views, opinions and/or findings expressed are those of the author(s) and should not be interpreted as representing the official views or policies of the Department of Defense or the U.S. Government. A.I. acknowledges support from the National Defense Science and Engineering (NDSEG) fellowship program. The authors acknowledge Matthew W. Puckett and Karl D. Nelson for initial contributions to the grating emitter photonic design. The authors thank Nan Yu and Sheng-Wey Chiow at NASA Jet Propulsion Laboratory, Joshua Hill of Army Research Lab, and Evan Salim and Chris Wood at ColdQuanta for useful discussions, Thorlabs Quantum Electronics for help in obtaining the booster amplifier, and Flame Feng and Alice Yu for help with the magnetic field coils and fiber coupling.

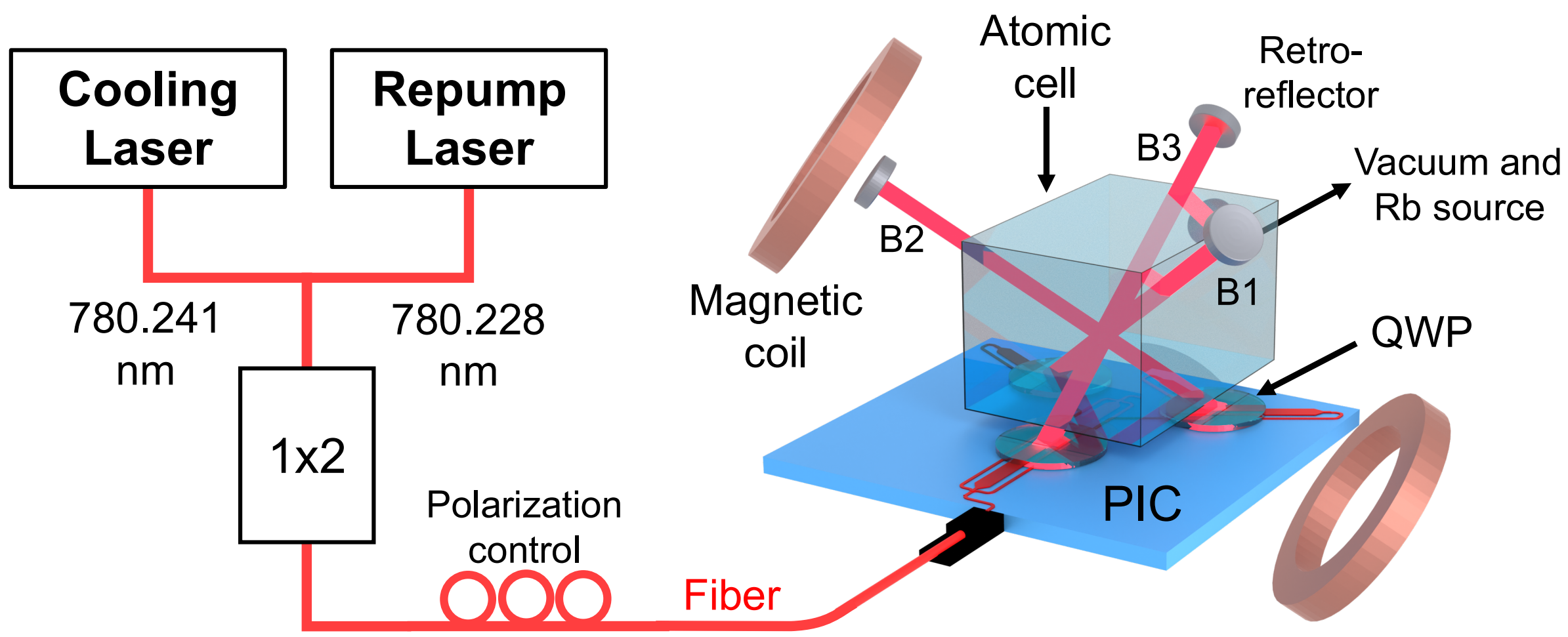

**Fig. 1. PICMOT system.** Laser cooling beam delivery PIC installed in the 3D MOT setup. The cooling and repump lasers are combined in a fiber coupler and connected to the PIC fiber input. The polarization is adjusted with a polarization controller. The quarter waveplate (QWP) holder layer is placed on the PIC and the stray-light filtering layer (not shown) is above the QWP holder. The PIC is placed under the glass cell containing the rubidium atoms and the magnetic field coils are aligned along beam B2 (the other beams are labeled B1 and B3). Retro-reflector mirror mounts are mounted above and used to overlap the counter-propagating beams.

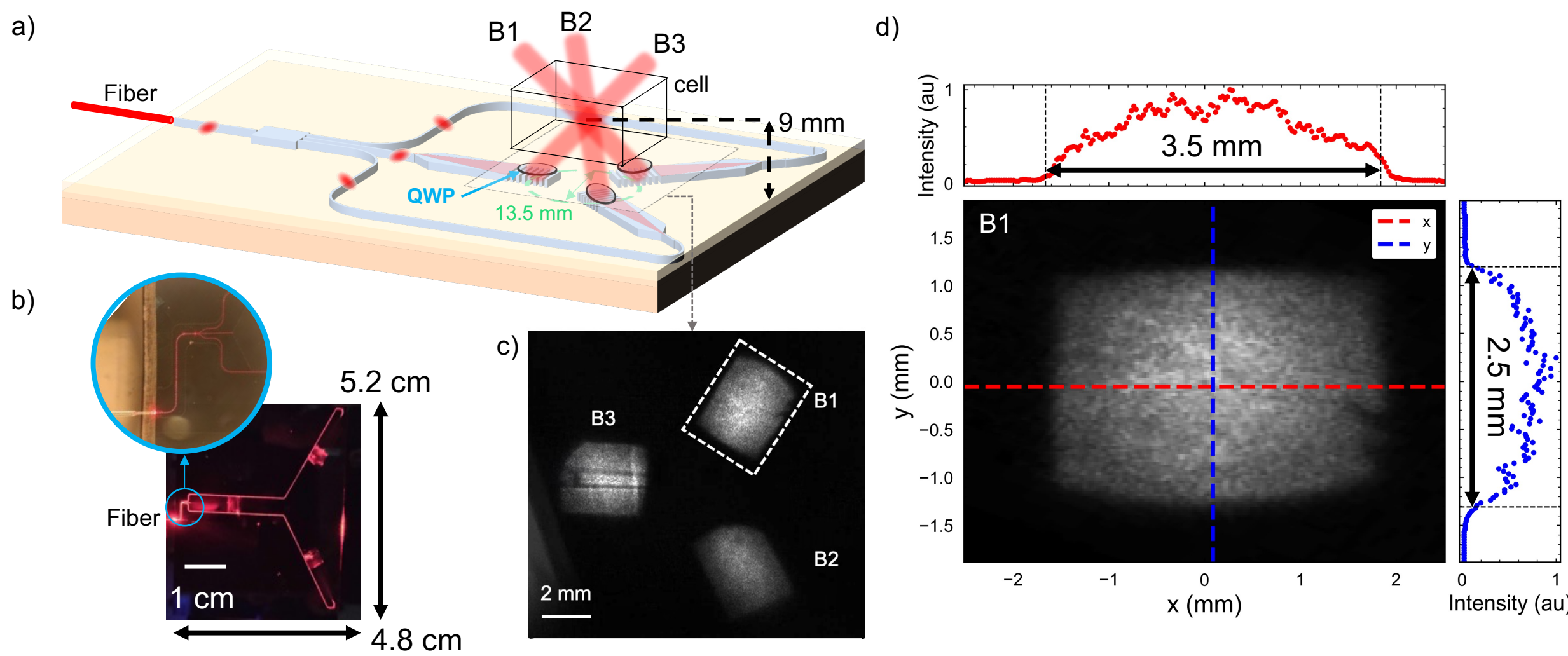

**Fig. 2. Laser cooling beam delivery PIC**. a) Layout of the laser cooling beam delivery PIC used to generate free-space beams for the 3D MOT. Fiber-coupled light is guided to a splitter and directed to the mode expanders and grating emitters. Bulk quarter waveplates are placed directly above each emitter. The dashed rectangle represents the height at which the beam profile is imaged. b) PIC illuminated with red laser light to demonstrate the light propagation. The inset shows the MMI splitter and fiber coupling. c) Image of the PIC beams incident on the paper sheet held parallel to the PIC surface at a height of 5 mm. The non-uniformity beam B3 is discussed in the Supplementary Note 4. d) Perpendicular cross-section image of beam B1 and the profiles of the beam intensity for two axes of the beam emitter and the corresponding $e^{-2}$ beam widths.

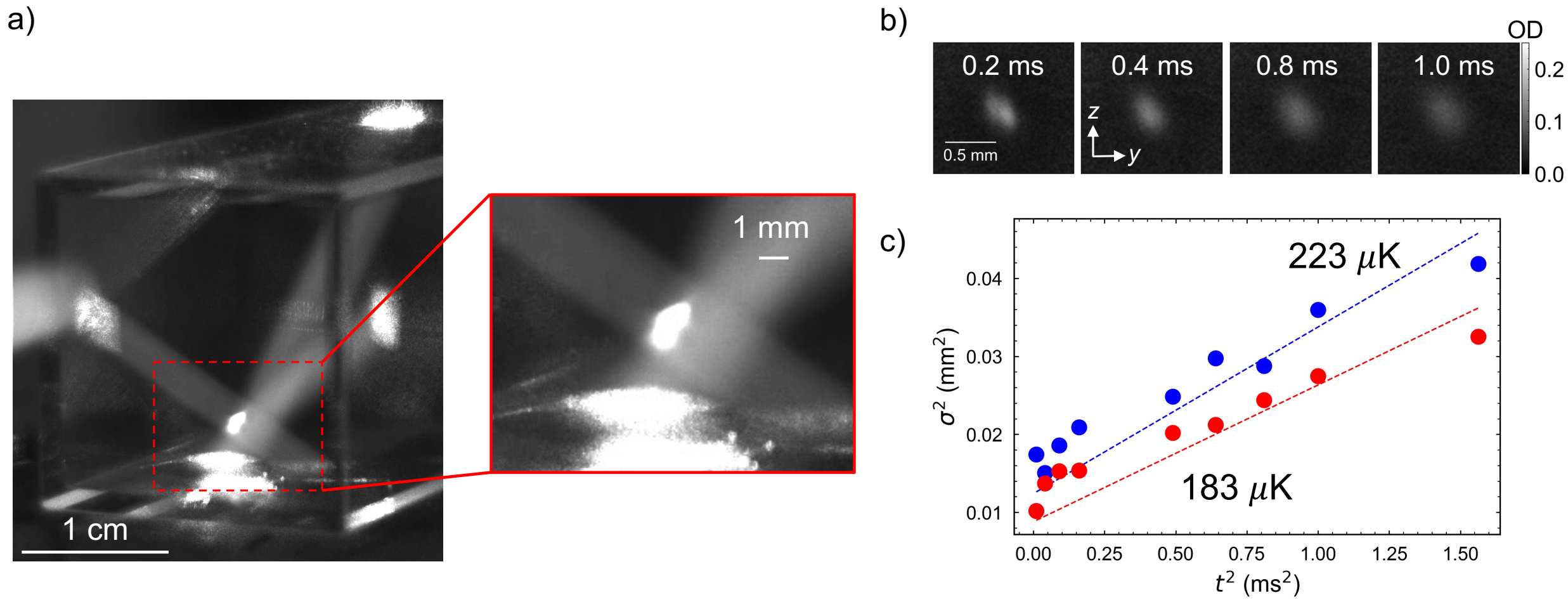

**Fig. 3. Rubidium 3D MOT demonstration**. a) View from a camera located on the side of the PICMOT setup. The side length of the glass cell is 20 mm. The trapping beams fluoresce as they propagate through the cell containing the atomic vapor. Inset: zoom-in of the MOT cloud (note some camera pixels are saturated). b) Averaged absorption images of the MOT cloud after free-expansion over several times. c) Squared cloud radius in the $z$ (blue) and $y$ (red) dimensions for squared TOF times $t^2$. The linear fits are for $\sigma_i^2 = \sigma_{i,0}^2 + \frac{k_B T_i}{m} t^2$, where $\sigma_i$ is the Gaussian standard deviation of the cloud along an axis $(y, z)$, $\sigma_{i,0}$ is the initial width of the cloud, $k_B$ is the Boltzmann constant, $m$ is the mass of a single $^{87}$Rb atom, and $T_i$ is the temperature along an axis.

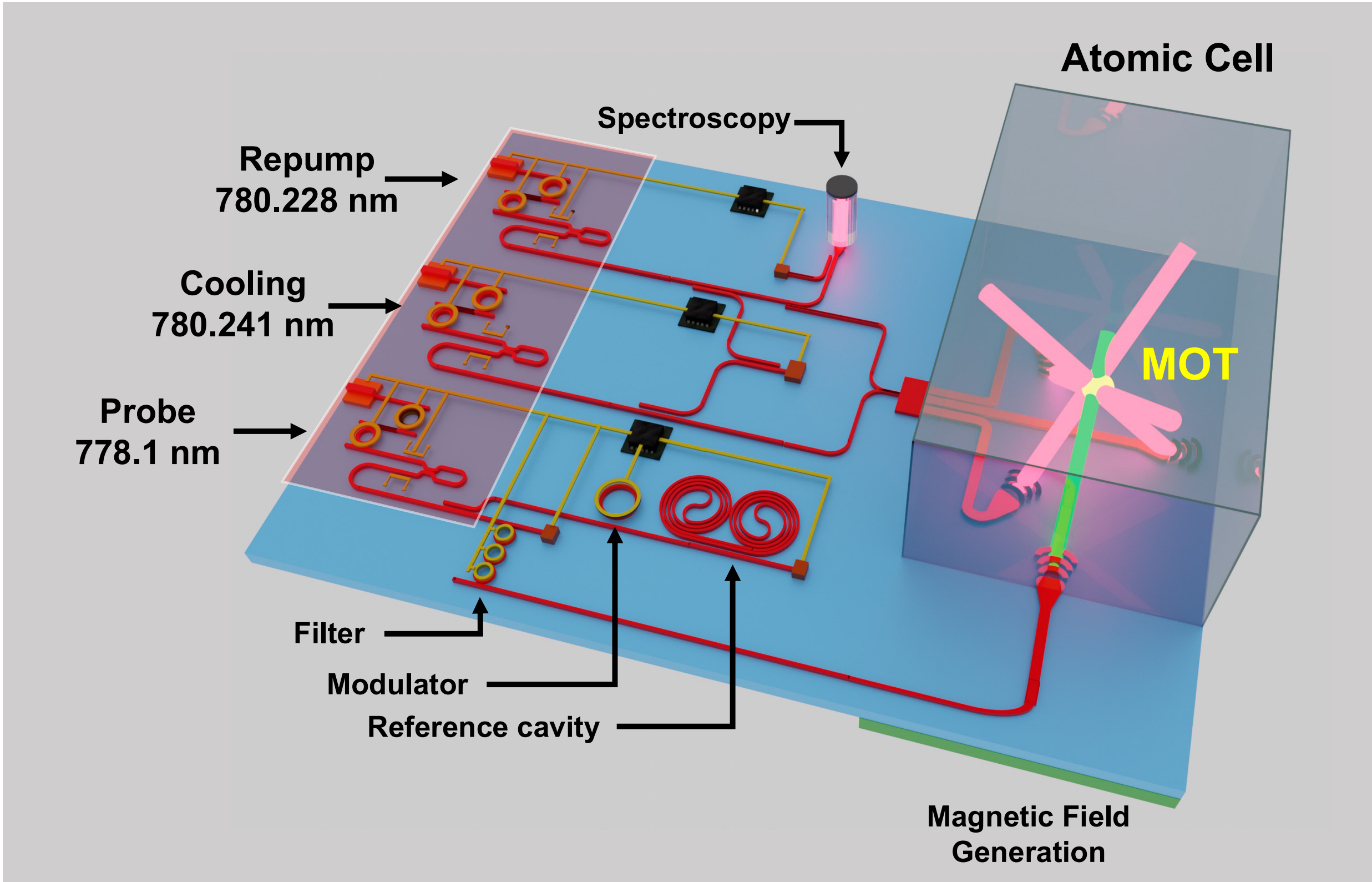

**Fig. 4. Example integration application of PIC to free-space beam delivery.** Illustrative example of how the large-area PIC beam emitters can be used in an integrated atomic-photonic system for cold atom trapping and probing. Heterogeneously integrated external cavity $Si_3N_4$ lasers[53] can be used to generate light for atom cooling, repumping, and probing. The lasers are each stabilized to a spectroscopy reference and a probe laser is locked to a reference cavity coil resonator for frequency noise reduction. PZT-actuated[52] modulators and shutters can be utilized for locking and probe light control. Each laser beam-line is delivered to the atomic species using the large-area photonic grating emitters of this work. The vacuum cell containing the atoms and the magnetic field generation layer can be planar integrated around the photonic chip.